\documentclass{iaus}
\usepackage{graphicx}

\title[Fluorescence and recombination in IC~418] 
{Excitation of emission lines by fluorescence and recombination in IC~418}

\author[V.~Escalante, C.~Morisset \& L.~Georgiev] 
{V.~Escalante$^1$, C.~Morisset$^2$
 \thanks{Temporarily at Instituto de Astrof\'{\i}sica de Canarias, 
38200, La Laguna, Tenerife, Spain.},
 \and L.~Georgiev$^2$}

\affiliation{$^1$CRyA, UNAM, Ap.~Postal 72--3, C.~P.~58091, 
Morelia, Michoac\'an, M\'exico \\[\affilskip]
$^2$Instituto de Astronom\'{\i}a, UNAM 
Ap.~Postal 70--264, C.~P.~04510, M\'exico, D.~F., M\'exico \\
emails: {\tt v.escalante@crya.unam.mx, chris.morisset@gmail.com, 
georgiev@astro.unam.mx}}

\pubyear{2011}
\volume{283}  
\pagerange{1--2}
\jname{Planetary Nebulae: An Eye to the Future}
\editors{A.~Manchado, L.~Stanghellini, \& D.~Sch\"oenberner, eds.}
\begin{document}

\maketitle

\begin{abstract}
We predict intensities of lines of CII, NI, NII,
OI and OII and compare them with a deep spectroscopic 
survey of IC~418 to test the effect of excitation of nebular 
emission lines by continuum fluorescence of starlight. 
Our calculations use a nebular model and 
a synthetic spectrum of its central star to 
take into account excitation of the lines by continuum 
fluorescence and recombination.
The NII spectrum is mostly produced by fluorescence 
due to the low excitation conditions of the nebula,
but many CII and OII lines have more excitation by 
fluorescence than recombination. 
In the neutral envelope, 
the NI permitted lines are excited by fluorescence,
and almost all the OI lines are excited by recombination.
Electron excitation produces the forbidden optical lines of OI,
but continuum fluorescence excites most of the NI forbidden line 
intensities.
Lines excited by fluorescence of light below the Lyman limit 
thus suggest a new diagnostic
to explore the photodissociation region of a nebula. 
\end{abstract}

\firstsection 

\section{Models}

\begin{figure}[t]
\begin{center}
 \includegraphics[width=2.6in]{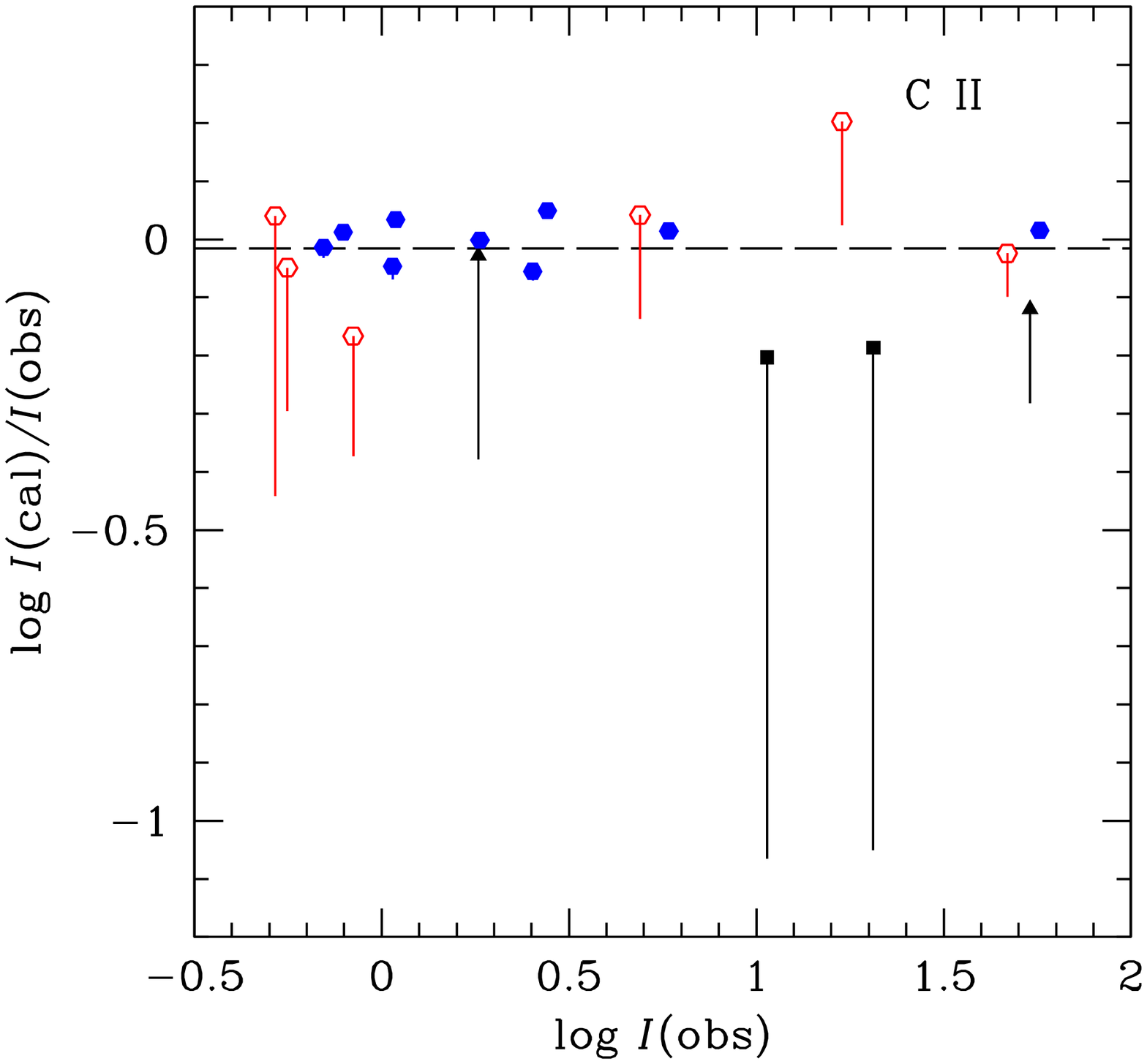} 
 \includegraphics[width=2.6in]{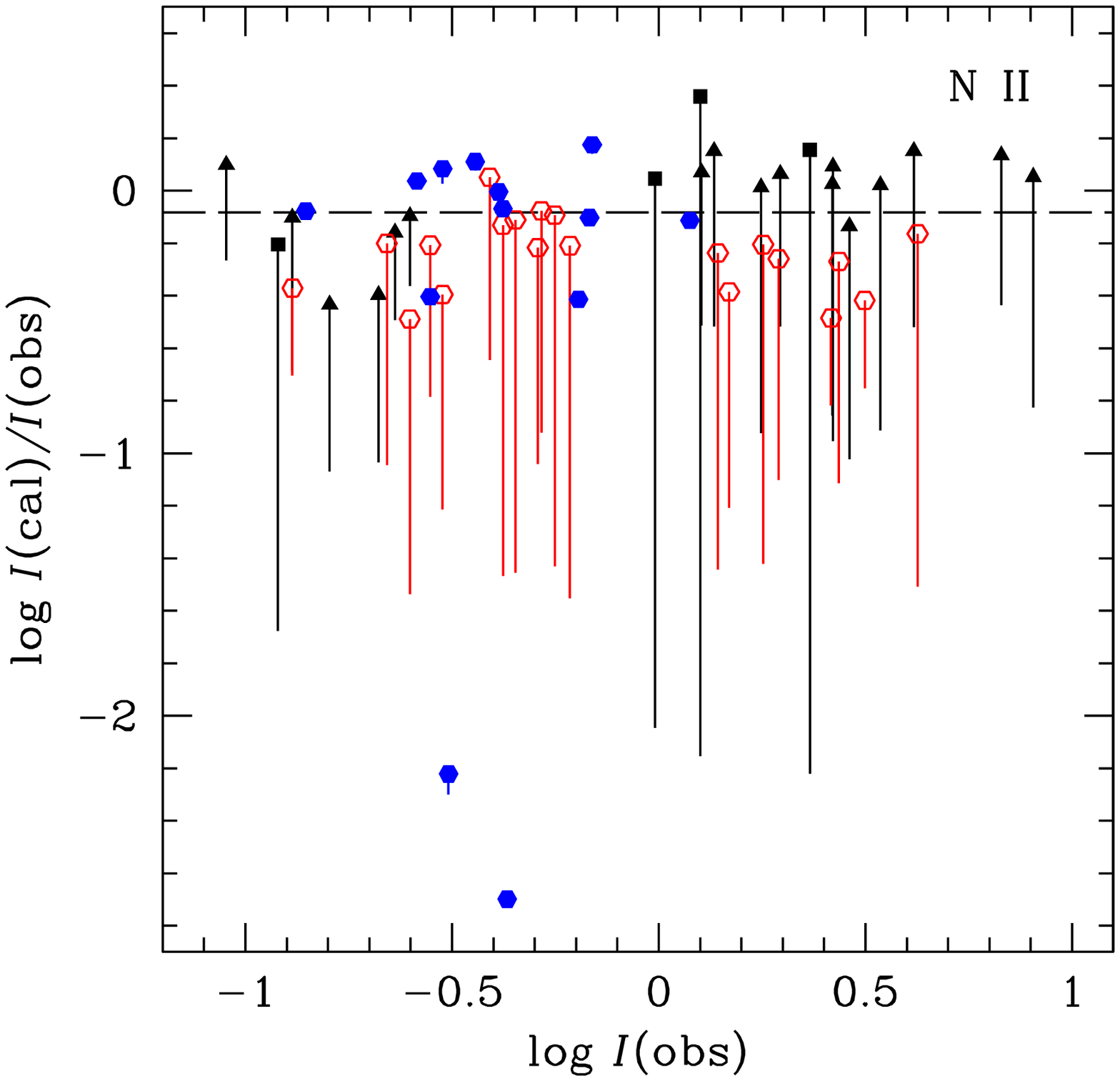} 
 \caption{Comparison of calculated and observed line intensities 
of C~II and N~II. 
Lines from s states: black squares, p states: black triangles, 
d states: red circles, f and g states: blue filled circles. The 
dashed line is the average $I({\rm calc})/I({\rm obs})$. Vertical 
lines show the increase of intensity due to fluorescence excitation 
(colour on line).}
   \label{fig1}
\end{center}
\end{figure}

\begin{figure}[t]
\begin{center}
 \includegraphics[width=2.6in]{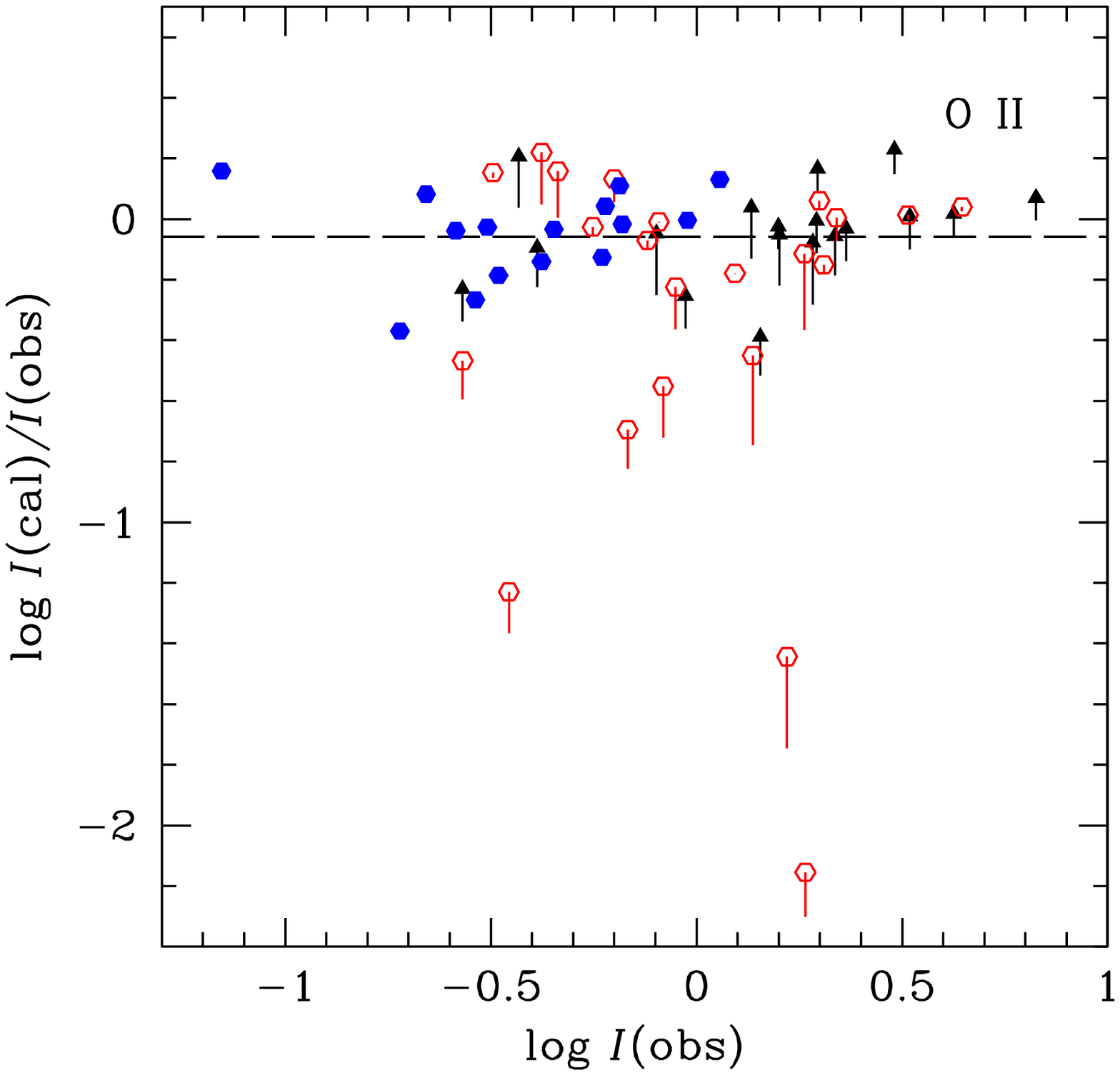} 
 \includegraphics[width=2.6in]{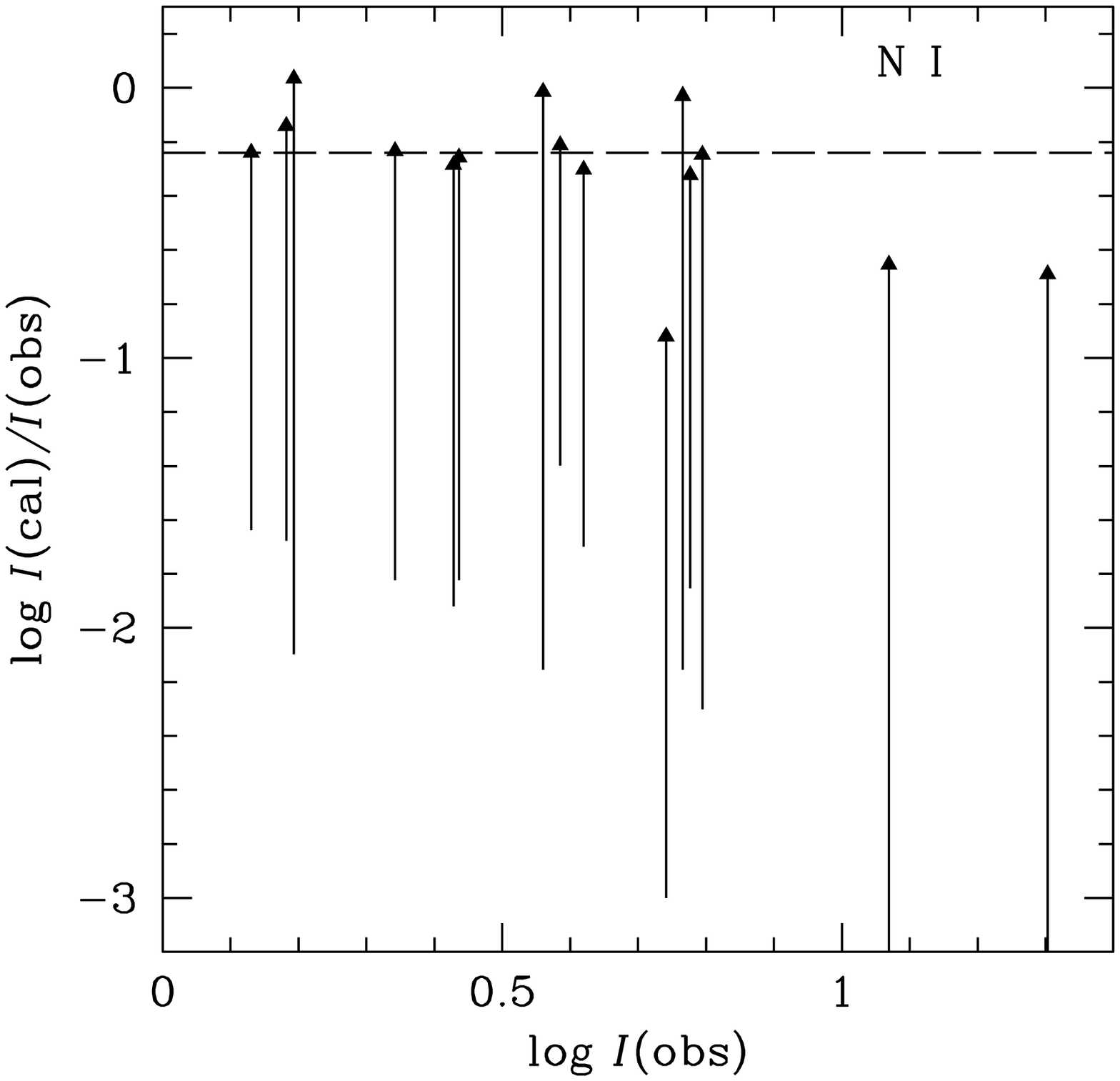} 
 \caption{Same as figure 1 for O~II and N~I (colour on line).}
   \label{fig2}
\end{center}
\end{figure}

\cite{chrisleonid} demonstrated that is it possible to 
construct consistent models of the central star atmosphere 
and the nebula to predict accurately the intensities of UV, 
optical and IR nebular and stellar emission lines in IC~418.  
We used similar models and included excitation of dipole--allowed 
transitions by fluorescence besides recombination 
to predict the intensities of 
189 emission lines of C~II, N~I, N~II, O~I, 
and O~II observed by \cite{sharpee}. 

The high--resolution synthetic spectrum produced by the 
CMFGEN code (\cite{cmfgen}) gives the 
opportunity to notice variations in the predicted intensities 
of some emission lines due to the expansion of the nebular gas. 
We used the CLOUDY nebular code (\cite{cloudy}) to 
model the nearly spherical nebula with a density profile 
similar to the one used by \cite{chrisleonid}. 
Fluorescence is important in the population of s, p, and 
some d states while recombination populates f and g states. 

\section{Results} 

The most important effect of fluorescence is in the N~II lines 
from the decay of s, p and most d states. Intense lines in 
planetary nebulae like $\rm3s^3P^o_1$--$\rm 3p^3D_4\lambda5666.63$ 
and $\rm3s^3P^o_2$--$\rm 3p^3D_3\lambda5679.56$ 
have fluorescence contributions of 78\% and 73\% respectively. 
C~II lines from most p and d states have between 20\% and 
69\% contribution by fluorescence. 
O~II lines are mostly excited by recombination, 
but fluorescence can contribute up to 50\% of the 
excitation of $\rm 3d^2F$ states (Figs.~\ref{fig1} 
and~\ref{fig2}). 

We can also predict some of the intensities of the N~I and 
O~I in a neutral shell around the ionized region. Most 
permitted O~I lines are produced by recombination, 
while N~I permitted and optical forbidden lines are 
excited by fluorescence (Fig.~\ref{fig2}).


\begin{thebibliography}{}

\bibitem[Escalante \& Morisset (2005)]{orion}
  {Escalante, V. \& Morisset, C.} 2005, 
  \textit{MNRAS}, 361, 813 

\bibitem[Ferland \etal\ (1998)]{cloudy}
  Ferland, G.~J., Korista, K.~T., Verner, D.~A., Ferguson, J.~W., 
  Kingdon, J.~B., Verner, E.~M., 1998, \textit{PASP}, 110, 761

\bibitem[Hillier \& Miller (1998)]{cmfgen}
  {Hillier, D.~J., Miller, D.~L.} 1998, \textit{ApJ}, 496, 407

\bibitem[Morisset \& Georgiev (2009)]{chrisleonid}
  {Morisset, C., \& Georgiev, L.} 2009, 
  \textit{A\&A}, 507, 1517 

\bibitem[Sharpee \etal\ (2003)]{sharpee}
  {Sharpee, B., Williams, R., Baldwin, J.A., van Hoof, P.A.M.} 2003, 
  \textit{ApJS}, 149, 157

\end{thebibliography}
\end{document}